\documentclass[letterpaper,twocolumn,american,prb,showpacs]{revtex4}
\usepackage[T1]{fontenc}
\usepackage[latin9]{inputenc}
\usepackage{bm}
\usepackage{amsmath}
\usepackage{graphicx}
\usepackage{esint}

\makeatletter
\@ifundefined{textcolor}{}
{%
 \definecolor{BLACK}{gray}{0}
 \definecolor{WHITE}{gray}{1}
 \definecolor{RED}{rgb}{1,0,0}
 \definecolor{GREEN}{rgb}{0,1,0}
 \definecolor{BLUE}{rgb}{0,0,1}
 \definecolor{CYAN}{cmyk}{1,0,0,0}
 \definecolor{MAGENTA}{cmyk}{0,1,0,0}
 \definecolor{YELLOW}{cmyk}{0,0,1,0}
 }

\makeatother

\usepackage{babel}

\begin{document}

\title{DMPK Equation for the Edge Transport of Quantum Spin Hall Insulator}

\author{Dafang Li and Junren Shi}

\affiliation{Institute of Physics, Chinese Academy of Sciences, Beijing 100190,
China}
\begin{abstract}
Using the random matrix theory, we investigate the ensemble statistics
of edge transport of a quantum spin Hall insulator with multiple edge
states in the presence of quenched disorder. Dorokhov-Mello-Pereyra-Kumar
equation applicable for such a system is established. It is found
that a two-dimensional quantum spin Hall insulator is effectively
a new type of one-dimensional (1D) quantum conductor with the different
ensemble statistics from that of the ordinary 1D quantum conductor
or the insulator with an even number of Kramers edge pairs. The ensemble
statistics provides a physical manifestation of the $Z_{2}$-classification
for the time-reversal invariant insulators.
\end{abstract}


\maketitle
One of the recent advances in condensed matter physics is the discovery
of the quantum spin Hall insulator (QSHI)~\cite{Kane2005a,Bernevig2006}.
QSHI is a new type of topological insulator, which is gaped in the
bulk, but has gapless edge modes that give rise to a quantized conductance.
The key theoretical observation is the $Z_{2}$-classification for
the time-reversal (TR) invariant insulating systems~\cite{Kane2005}:
a two-dimensional (2D) insulator with an odd number of Kramers pairs
of edge states and that with an even number are topologically distinct,
and the QSHI has an odd number of Kramers pairs at its edge. Such
a classification has been established by the analyses on the topological
structure of the Bloch bands~\cite{Kane2005,Moore2007}, and its
robustness against the imperfections, such as the electron-electron
interaction~\cite{Wu2006} and disorders~\cite{Xu2006,Obuse2007,Obuse2008},
has also been discussed. Experimentally, a quantized conductance is
observed in HgTe quantum wells, and is taken as the signature of the
QSHI phase~\cite{Konig2007}, albeit not conclusively. Other experimental
techniques, such as ARPES, are also employed for searching the new
QSHIs~\cite{Hsieh2008}. At present, it is highly desirable to have
more associations between the abstract $Z_{2}$-classification and
the physically measurable properties.

In this Rapid Communication, we investigate the ensemble statistics
of the edge transport of QSHI in the presence of quenched disorder.
In essence, a two-dimensional (2D) QSHI is effectively a one-dimensional
(1D) quantum conductor with an odd number of Kramers pairs of conducting
channels. Such a 1D quantum conductor is actually a new species that
can only be realized at the edge of a 2D QSHI~\cite{Wu2006}, different
from the ordinary 1D conductors which always have an even number of
Kramers pairs of conducting channels. We establish the Dorokhov-Mello-Pereyra-Kumar
(DMPK) equation~\cite{Beenakker1997} applicable for such a system,
upon which the ensemble statistics of the edge transport of the QSHI
is investigated. The distinct ensemble statistics of the edge transport
of the QSHI presents a physical manifestation of the $Z_{2}$-classification,
and could be a useful probe for identifying the new TR invariant topological
insulators. 

\begin{figure}
\includegraphics[width=1\columnwidth]{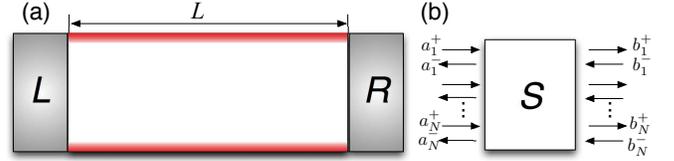}

\caption{\label{fig:structure} (color online) (a) The geometry of the system:
an insulator with multiple edge conducting channels (represented by
the color shaded bands at the two edges) is connected to the left
and right measurement leads. (b) The transport along each of edges
can be characterized by a $S$-matrix. $a_{i}^{+}$ ($b_{i}^{+}$)
and $a_{i}^{-}$ ($b_{i}^{-}$) denote the right-going and left-going
wave amplitudes, respectively. $a_{i}^{\pm}$ (or $b_{i}^{\pm}$)
with the same index $i$ are related by the TR and form a Kramers
pair. $N$ denotes the total number of Kramers pairs at each edge,
and can be odd (QSHI) or even (ordinary insulator). }

\end{figure}

We consider a configuration shown in Fig.~\ref{fig:structure}(a).
Because the insulating bulk prevents the direct communication between
the two edges, the system can be considered effectively as two independent
1D quantum conductors arranged in parallel. Each 1D conductor has
$N$ Kramers pairs of conducting channels. We assume that the spin-orbit
coupling is present, so the spins are not conserved in general. We
do not assume the origin of the edge modes: they can be a result of
the topological structure of the bulk bands, or from the extrinsic
origins such as the surface dangling bonds. 

In general, the transmission along the 1D conductor can be characterized
by a $2N\times2N$ $S$-matrix, which relates the incoming ($\psi_{\mathrm{in}}$)
and outgoing ($\psi_{\mathrm{out}}$) wave amplitudes:\begin{equation}
\psi_{\mathrm{out}}=S\psi_{\mathrm{in}}\end{equation}
where $\psi_{\mathrm{in}}\equiv(a_{1}^{+},a_{2}^{+}\dots a_{N}^{+};b_{1}^{-},b_{2}^{-}\dots b_{N}^{-})^{T}$
and $\psi_{\mathrm{out}}\equiv(a_{1}^{-},a_{2}^{-}\dots a_{N}^{-};b_{1}^{+},b_{2}^{+}\dots b_{N}^{+})^{T}$
(see Fig.~\ref{fig:structure}(b)). In our labeling of the channel
numbers, TR symmetry imposes the constraint on the $S$-matrix~\cite{Bardarson2008}:\begin{equation}
S^{T}=-S,\end{equation}
Moreover, the current conserving implies $S$-matrix must be unitary:
$S^{\dagger}S=I$.

Under these constraints, the polar decomposition of the $S$-matrix
reads~\cite{Beenakker1997,Bardarson2008}:\begin{equation}
S=\left[\begin{array}{cc}
U^{T} & 0\\
0 & V^{T}\end{array}\right]\left[\begin{array}{cc}
\Sigma & \mathcal{T}\\
-\mathcal{T} & -\Sigma\end{array}\right]\left[\begin{array}{cc}
U & 0\\
0 & V\end{array}\right]\end{equation}
where $U$ and $V$ are $N\times N$ unitary matrices, $\Sigma$ is
a block diagonal matrix $\Sigma\equiv\Sigma_{1}\oplus\Sigma_{2}\oplus\cdots\oplus\Sigma_{n}(\oplus\bm{0}_{1\times1})$
with $\Sigma_{i}=\sqrt{1-T_{i}}\left[\begin{array}{cc}
0 & 1\\
-1 & 0\end{array}\right]$, and $\mathcal{T}=\mathrm{diag}[\sqrt{T_{1}},\sqrt{T_{1}},\sqrt{T_{2}},\sqrt{T_{2}}\cdots\sqrt{T_{n}},\sqrt{T_{n}}(,1)]$,
for the even $N\equiv2n$ (odd $N\equiv2n+1$), $T_{i}$ denotes the
$i$th transmission eigenvalue. One immediately sees that for the
odd $N$, there is always one conducting channel that has the perfect
transmission, without being adversely affected by the disorder. This
is the reason behind the robust edge transport of QSHI. 

The different ensemble statistics for the 1D conductors with the odd
and the even $N$ can already be observed if we compute the volume
element of the configuration space expanded by the independent parameters
of the $S$-matrix: $\mathrm{d}\mu(S)=J_{0}\mathrm{d}\mu(U)\mathrm{d}\mu(V)\prod_{i}\mathrm{d}T_{i}$.
To get the invariant measure $J_{0}$, we calculate $\mathrm{d}s^{2}=\mathrm{Tr}[\mathrm{d}S^{\dagger}\mathrm{d}S]\equiv g_{ij}\mathrm{d}x_{i}\mathrm{d}x_{j}$,
and $J_{0}=\sqrt{\mathrm{det}[g_{ij}]}$, where $x\equiv\{U_{ij},V_{ij},T_{i}\}$~\cite{Mello1988}.
We obtain:\begin{multline}
J_{0}=\prod_{i<j=1}^{n}(T_{i}-T_{j})^{4}\begin{cases}
\prod_{i=1}^{n}T_{i}, & \mathrm{even\,}N\\
\prod_{i=1}^{n}T_{i}(1-T_{i})^{2}, & \mathrm{odd\,}N\end{cases}.\label{eq:measure}\end{multline}

To determine the ensemble statistics of the 1D conductor, we derive
the Dorokhov-Mello-Pereyra-Kumar (DMPK) equation for evolution of
the joint probability distribution function of the transmission eigenvalues
about the length $L$: $P(T_{1},T_{2}\cdots T_{n};L)$~\cite{Beenakker1997}.
Basically, we consider a 1D quantum conductor of length $L$, and
compute the change of the transmission eigenvalues upon attachment
of a thin slice of length $\delta L$. Using the perturbation approach,
we obtain~\cite{Mello1988,Beenakker1997}:\begin{align}
\frac{1}{\delta s}\left\langle \delta T_{i}\right\rangle  & =-T_{i}+\frac{2n}{N(N-1)}T_{i}\nonumber \\
 & \times\left(1-T_{i}+2\sum_{i\ne j}\frac{T_{i}+T_{j}-2T_{i}T_{j}}{T_{i}-T_{j}}\right),\\
\frac{1}{\delta s}\left\langle \delta T_{i}\delta T_{j}\right\rangle  & =\frac{4n}{N(N-1)}T_{i}^{2}\left(1-T_{i}\right)\delta_{ij},\end{align}
where $\delta s\equiv\delta L/l$, $l$ is the mean free path defined
by the first moment of the transmission eigenvalues of the thin slice:
$\delta L/l=1-\left\langle \sum_{i=1}^{n}T_{i}(\delta L)\right\rangle /n$.
The third and higher moments vanish at the first order of $\delta s$.
DMPK equation is just the Fokker-Planker equation for the evolution
of the distribution function $P$: \begin{equation}
\frac{\partial P_{\lambda}}{\partial s}=\frac{2}{\gamma_{N}}\sum_{i=1}^{n}\frac{\partial}{\partial\lambda_{i}}\left\{ \lambda_{i}(1+\lambda_{i})J\frac{\partial}{\partial\lambda_{i}}\left(\frac{P_{\lambda}}{J}\right)\right\} ,\label{eq:DMPK}\end{equation}
where $\gamma_{N}\equiv N(N-1)/n$, and we have re-expressed the distribution
function in a new set of variables $\lambda_{i}\equiv(1-T_{i})/T_{i}$,
$P_{\lambda}(\lambda_{1},\lambda_{2}\cdots\lambda_{n},s)\equiv P(T_{1},T_{2},\cdots T_{n},L)\prod_{i=1}^{n}(1+\lambda_{i})^{-2}$,
and\begin{equation}
J=\prod_{i<j=1}^{n}(\lambda_{i}-\lambda_{j})^{4}\times\begin{cases}
1, & \mbox{even}\, N\\
\prod_{i=1}^{n}\lambda_{i}^{2}, & \mbox{odd}\, N\end{cases},\label{eq:J}\end{equation}
which actually coincides with Eq.~(\ref{eq:measure}) except for
an unimportant denominator. We note that a similar DMPK equation was
derived by Takane for the metallic carbon nanotubes~\cite{Takane2004}.

Equations (\ref{eq:DMPK}--\ref{eq:J}) are the central result of
this paper. The equation is reduced to the usual DMPK equation of
the ordinary 1D conductor of the symplectic ensemble ($\beta=4$)
for the even $N\equiv2n$~\cite{Beenakker1997}. On the other hand,
for the odd $N\equiv2n+1$, the equation is modified, and one expects
a different distribution of the transmission eigenvalues. We will
spell out its implications in the following.

Equation (\ref{eq:DMPK}) turns out to be completely integrable for
both even~\cite{Caselle1995} and odd $N$. To see this, we adopt
a new set of variables $\{x_{i}\}$ that are related to $\{\lambda_{i}\}$
by $\lambda_{i}=\sinh^{2}x_{i}$, and $P_{x}(\{x_{i}\},s)\equiv P_{\lambda}\prod_{i=1}^{n}\sinh2x_{i}$.
We further make the substitution $P_{x}=\xi^{2}(x)\Psi(x,s)$, with\begin{multline}
\xi(x)=\prod_{i<j=1}^{n}\sinh^{2}(x_{i}+x_{j})\sinh^{2}(x_{j}-x_{j})\\
\times\prod_{i=1}^{n}\sinh^{1/2}2x_{i}\begin{cases}
1, & \mbox{even}\, N\\
\sinh^{2}x_{i} & \mbox{odd}\, N\end{cases},\label{eq:xi}\end{multline}
and the equation is transformed to:\begin{equation}
-\frac{\partial\Psi}{\partial s}=-\frac{1}{2\gamma_{N}}\left[\sum_{i=1}^{n}\frac{1}{\xi(x)^{2}}\frac{\partial}{\partial x_{i}}\xi(x)^{2}\frac{\partial}{\partial x_{i}}\right]\Psi.\label{eq:Psi}\end{equation}
We can then identify the operator inside the square bracket of the
rhs. of Eq.~(\ref{eq:Psi}) being the radial part of the Laplace-Beltrami
operator for the irreducible symmetric space $\mathrm{SO}^{*}(2N)/U(N)$~\cite{Caselle1995,Olshanetsky1983}.
In particular, for the odd $N$, Equation (\ref{eq:xi}) corresponds
to a root system $BC_{n}$ of $\bm{\alpha}\equiv\{\pm\bm{e}_{i},\pm2\bm{e}_{i},\pm\bm{e}_{i}\pm\bm{e}_{j}\}$,
and has the appropriate multiplicity $m_{\bm{\alpha}}=\{4,1,4\}$
for a successful mapping to the Laplace-Beltrami operator (see Table
B1 of Ref.~\onlinecite{Olshanetsky1983}). This allows us to express
the distribution function $P_{x}(x,s)$ as a superposition of the
zonal spherical functions $\Phi_{k}(x)$~\cite{Caselle1995}:\begin{equation}
P_{x}(\{x\},s)=C(s)\xi^{2}(x)\int\Phi_{k}(x)e^{-k^{2}s/2\gamma_{N}}\frac{\mathrm{d}^{n}k}{|c(k)|^{2}}.\label{eq:expansion}\end{equation}
For the odd $N$, the Gindikin-Karpelevich formula (Eq.~(C12) of
Ref.~\onlinecite{Olshanetsky1983}) yields:\begin{multline}
c(k)=\prod_{i<j=1}^{n}\frac{1}{g\left(\frac{k_{i}+k_{j}}{2}\right)g\left(\frac{k_{i}+k_{j}}{2}\right)}\\
\times\prod_{i=1}^{n}\frac{1}{(1+ik_{i})^{2}}\frac{\Gamma\left(i\frac{k_{i}}{2}\right)}{\Gamma\left(\frac{1}{2}+i\frac{k_{i}}{2}\right)},\end{multline}
where $g(x)\equiv\mathrm{i}x(1+\mathrm{i}x)$. The zonal spherical
function $\Phi_{k}(x)$ can be constructed by a recurrent procedure
(Eqs.~(8.7-8.10) of Ref.~\onlinecite{Olshanetsky1983}). 

Using the asymptotic expansion of $\Phi_{k}(x)$ and following the
same line of derivations as shown in Ref.~\onlinecite{Caselle1995},
we can determine the asymptotic forms the distribution function. In
the localization regime ($s\gg N$):\begin{equation}
P_{x}(\{x_{i}\},s)\propto\prod_{i=1}^{n}\exp\left[-(\gamma_{N}/2s)(x_{i}-\bar{x}_{i})^{2}\right],\label{eq:P localization}\end{equation}
where $\bar{x}_{i}=(s/\gamma_{N})[3+4(i-1)]$. It follows that the
average conductance $\sigma\sim1+2\exp(-L/2\xi)$ with the localization
length $\xi=Nl/2$ for the odd $N$, compared with $\xi=2(N-1)l$
for the ordinary 1D conductor (even $N$)~\cite{Beenakker1997}.
The conductance will have a log-normal distribution in this regime.

In the diffusive regime ($1\ll s\ll N$):\begin{multline}
P_{x}(\{x_{i}\},s)\propto\prod_{i<j}\left(\sinh^{2}x_{i}-\sinh^{2}x_{j}\right)^{2}\left(x_{i}^{2}-x_{j}^{2}\right){}^{2}\\
\times\prod_{i}e^{-x_{i}^{2}\gamma_{N}/2s}\left(x_{i}\sinh2x_{i}\right)^{1/2}(x_{i}\sinh x_{i})^{2}.\label{eq:P diffusive}\end{multline}
Compared with the ordinary 1D conductor~\cite{Caselle1995,Muttalib2005},
the distribution function acquires an extra factor $\prod_{i}(x_{i}\sinh x_{i})^{2}$.
Note that the correlations between the different transmission eigenvalues
do not change.

We can determine the average and variance of the conductance in the
regime $s\ll N$ using the method of moments of Mello and Stone~\cite{Mello1991,Beenakker1997},
which computes the moments of $M_{q}\equiv\sum_{i=1}^{n}T_{i}^{q}$
as expansion in inverse powers of $N$. From the DMPK equation (\ref{eq:DMPK}),
we can establish a chain of the coupled evolution equations for moments
of $M_{q}$:\begin{multline}
\frac{\gamma_{N}}{2}\frac{\partial\left\langle M_{q}^{p}\right\rangle }{\partial s}=2pq\sum_{k=1}^{q-1}\left\langle M_{q}^{p-1}M_{k}(M_{q-k}-M_{q-k+1})\right\rangle \\
-2pq\left\langle M_{q}^{p}M_{1}\right\rangle +pq^{2}\left\langle M_{q}^{p-1}M_{q+1}\right\rangle -pq\left[q\mp1\right]\left\langle M_{q}^{p}\right\rangle \\
+p(p-1)q^{2}\left\langle M_{q}^{p-2}\left(M_{2q}-M_{2q+1}\right)\right\rangle ,\end{multline}
where $\mp$ sign stands for the odd ($+$) and even ($-$) $N$,
respectively. Since $M_{q}^{p}=\mathcal{O}(N^{p})$ in the particular
regime, we can close the above equation order by order in the large
$N$ limit. Noting that the average conductance $\sigma/\sigma_{0}=2\left\langle M_{1}\right\rangle +Z_{2}$
and the variance $\mathrm{var}(\sigma)/\sigma_{0}^{2}=4\left(\left\langle M_{1}^{2}\right\rangle -\left\langle M_{1}\right\rangle ^{2}\right)$,
$\sigma_{0}\equiv e^{2}/h$, we obtain:\begin{align}
\frac{\delta\sigma}{\sigma_{0}} & =\frac{s^{3}}{3(1+s)^{3}}+Z_{2}\frac{s}{(1+s)^{2}}+\mathcal{O}\left(N^{-1}\right),\label{eq:deltas}\\
\frac{\mathrm{var}(\sigma)}{\sigma_{0}^{2}} & =\frac{2}{15}\left[1-\frac{1+6s}{(1+s)^{6}}\right]+\mathcal{O}\left(N^{-1}\right),\label{eq:vars}\end{align}
where $\delta\sigma\equiv\sigma-N\sigma_{0}/(1+s)$ is the weak localization
correction to the conductance, and we have introduced an index $Z_{2}$
that takes the value of $0$ (1) for the even (odd) $N$. Compared
with the ordinary 1D conductor, the edge transport of QSHI will have
a different weak localization correction but the same universal conductance
fluctuation. 

\begin{figure}
\includegraphics[width=1\columnwidth]{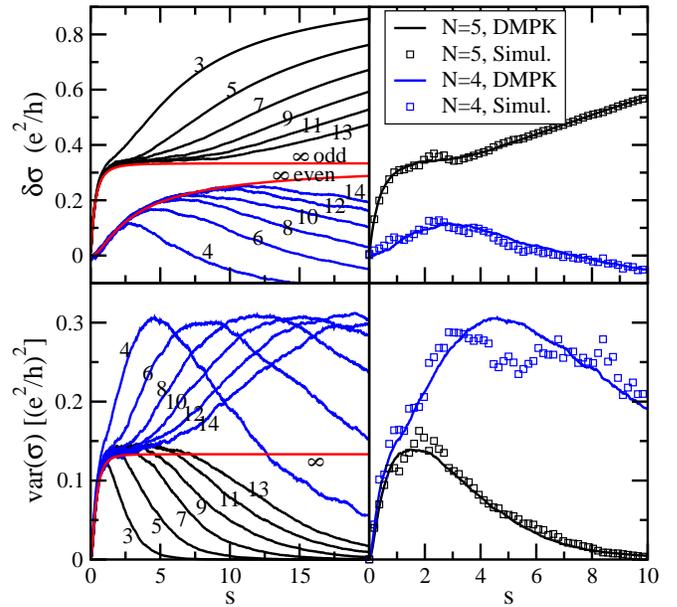}

\caption{\label{fig:ds-vars} (color online) The weak localization correction
to the conductance ($\delta\sigma$) and the variance of conductance
($\mathrm{var}(\sigma)$).$ $ Left: The results obtained from the
numerical solutions of DMPK equation for different values of $N$
(the numbers by the curves). The $N\rightarrow\infty$ asymptotic
behaviors Eqs.~(\ref{eq:deltas}--\ref{eq:vars}) are indicated by
the red curves. Right: The comparison between the results from the
DMPK equation and that from the simulation of the 2D model Eq.~(\ref{eq:KM}).
The results for $N=4$ and $N=5$ are shown. In the 2D simulation,
the parameters for the Kane-Mele model are: $t=1$, $\lambda_{\mathrm{so}}=0.2$,
$\lambda_{R}=0.2$, $\lambda_{v}=0$ (see Ref.~\onlinecite{Kane2005}
for the definitions of these parameters), and the other parameters
are: $W=0.6$, $t_{1}=0.2$. The stripe has a width 48 lattice sites
(LS) along the zig-zag direction. We have determined the mean free
paths to be $l=2240$ LS ($N=4$) and $l=1680$ LS ($N=5$), respectively.
200 disorder configurations are averaged.}

\end{figure}

We have numerically solved the DMPK equation for different values
of $N$, using a Monte-Carlo approach that simulates the diffusion
of the transmission eigenvalues. The weak localization correction
and the fluctuation of the conductance are calculated, shown in the
left panel of Fig.~\ref{fig:ds-vars}. Equations (\ref{eq:deltas})--(\ref{eq:vars})
well predict the behaviors for both $\delta\sigma$ and $\mathrm{var}(\sigma)$
in the regime $s\ll N$. The difference between the odd $N$ and the
even $N$ is evident. It is interesting to observe that although the
variances of the conductance have the same asymptotic behaviors in
the regime $s\ll N$ for the odd and even $N$ (see Eq.~(\ref{eq:vars}),
they are very different in the crossover regime ($s\sim N$): $\mathrm{var}(\sigma)$
rapidly decreases to zero for the odd $N$, while it continues to
increase and peaks at a value of $0.3(e^{2}/h)^{2}$ for the even
$N$. 

We further test our results against a real model of the 2D QSHI. We
consider a system of a stack of $N$ layers of honeycomb lattice,
each of which is described by the tight-binding Hamiltonian introduced
by Kane and Mele~\cite{Kane2005}:\begin{multline}
H=\sum_{l=1}^{N}H_{l}^{\mathrm{KM}}+\sum_{il}\epsilon_{il}c_{il}^{\dagger}c_{il}+t_{1}\sum_{\left\langle ll^{\prime}\right\rangle }c_{il}^{\dagger}c_{il^{\prime}},\label{eq:KM}\end{multline}
where $c_{il}$ ($c_{il}^{\dagger}$) denotes the annihilation (creation)
operator for lattice site $i$ at the $l$-th layer, $H_{l}^{\mathrm{KM}}$
denotes the Hamiltonian for the $l$th-layer and has the same form
as Eq.~(1) of Ref.~\onlinecite{Kane2005}, $\epsilon_{il}$ is the
random site energy that uniformly distributes in $[-W/2,\, W/2]$,
and the last term introduces the hopping between the neighboring layers
with a hopping constant $t_{1}$. With the appropriate parameters,
the system becomes an insulator with $N$-pairs of edge states. The
presence of the Rashba spin-orbit coupling in $H^{\mathrm{KM}}$ and
the disordered site energies will introduce backscattering between
different edge channels. In the simulation, a small value of $W$
is chosen, so that the bulk is still insulating. This is different
from the previous numerical investigations which concern more on the
annihilation of the edge states by the strong disorder due to the
breakdown of the bulk gap~\cite{Onoda2007,Qiao2008}. We have adopted
an iterative approach based on the non-equilibrium Green's function
to calculate the conductance of a stripe of varying length~\cite{Kazymyrenko2008}.
The results are presented in the right panel of Fig.~\ref{fig:ds-vars}.
It is evident that both the weak localization correction to the conductance
and the variance fit well with those predicted from the DMPK equation.
It justifies our presumption that a 2D QSHI is effectively a 1D quantum
conductor, and can be described by the DMPK equation (\ref{eq:DMPK}--\ref{eq:J}).

\begin{figure}
\includegraphics[width=1\columnwidth]{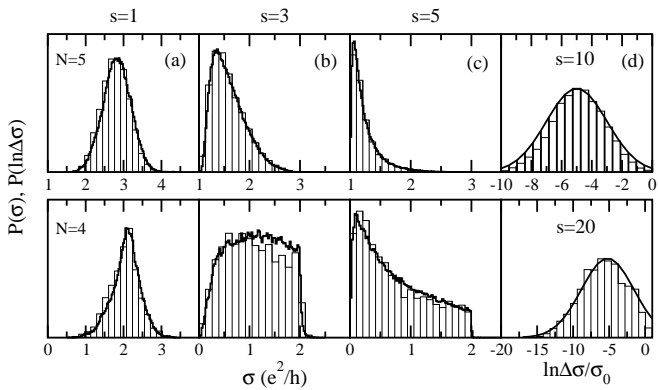}

\caption{\label{fig:distribution}The distribution of conductance for $N=4$
(lower) and $5$ (upper). The drop lines shows the results from the
DMPK equation. The solid lines indicates the distributions predicted
by the asymptotic formulae Eq.~(\ref{eq:P diffusive}) {[}(a)--(c){]}
and (\ref{eq:P localization}) {[}(d){]}. $\Delta\sigma\equiv\sigma(s)-\sigma(\infty)$.}

\end{figure}

Figure~\ref{fig:distribution} shows the distributions of conductance
for different values of $s$. A crossover from the Gaussian distribution
at the ballistic limit ($s=1$) to the log-normal distribution in
the localization regime ($s=10,\,20$) can be observed. The difference
between the odd and the even $N$ is the most notable in the crossover
regime ($s=3,\,5$), where the distribution for $N=5$ shows a smooth
high conductance tail, while that for $N=4$ has the a sharp threshold. 

Finally, we discuss the possible experimental verification of our
theoretical predictions. A 2D QSHI with the multiple pairs of edge
states can be realized by confining the 3D QSHI in one direction~\cite{Fu2007a}.
An alternative and more flexible way is to put an ordinary mesoscopic
1D quantum wire in the proximity of the edge of a 2D QSHI sample,
and couple them through a controllable gate. The ensemble statistics
can then be measured for the combined system. By varying the coupling
strength between the 1D wire and 2D QSHI, one expects a crossover
of the ensemble statistics from the even $N$ to the odd $N$. 

We thank C. Murdy for useful information. This work is supported by
NSF of China (No.~10604063 and 10734110) and the National Basic Research
Program of China (No.~2009CB929101 and 2006CB921304).

\bibliographystyle{apsrev}
\bibliography{/Users/shi/Documents/Papers/Papers}

\end{document}